\documentclass[preprint,11pt]{aastex}

\begin{document}

\title{The Origin of Planetary Impactors in the Inner Solar System\vspace*{0.1in}}

\author{Robert G.~Strom\altaffilmark{1},
Renu Malhotra\altaffilmark{1},
Takashi Ito\altaffilmark{2},
Fumi Yoshida\altaffilmark{2}, and
David A.~Kring\altaffilmark{1}}

\altaffiltext{1}{Lunar and Planetary Laboratory, University of Arizona, Tucson, AZ 85721, USA.}

\altaffiltext{2}{National Astronomical Observatory, Osawa, Mitaka, Tokyo 181-8588, Japan.  \\
\\
``This manuscript has been published in Science (September 16, 2005).  This
version pre-dates the final editing.  Please refer to the complete
version of record at {\tt http://www.sciencemag.org/}.  This manuscript may
not be reproduced or used in any manner that does not fall within the
fair use provisions of the Copyright Act without the prior, written
permission of AAAS."\\
Correspondence should be addressed to Renu Malhotra, email: renu@lpl.arizona.edu
}


\begin{abstract}
New insights into the history of the inner solar system
are derived from the impact cratering record of the Moon,
Mars, Venus and Mercury, and from the size distributions
of asteroid populations. Old craters from a unique  
period of heavy bombardment that ended $\sim$3.8 billion years
ago were made by asteroids that were dynamically ejected
from the main asteroid belt, possibly due to the orbital
migration of the giant planets.  The impactors of the
past $\sim$3.8 billion years have a size distribution quite
different from the main belt asteroids, but very similar
to the population of near-Earth asteroids.
\end{abstract}

The Moon and all the terrestrial planets were resurfaced during 
a period of intense impact cratering that occurred between the 
time of their accretion, $\sim\!4.5$ Ga, and $\sim\!3.85$ Ga. 
The lunar cratering record and the radiometrically dated Apollo 
samples have shown that the intense bombardment of the Moon ended 
at $\sim\!3.85$ Ga; the impact flux since that time to the present has 
been at least an order of magnitude smaller. The 3.85 Ga epoch 
might represent the final end of an era of steadily declining 
large impacts (the tail end of the accretion of the planets). 
However, it has also been argued that only a sudden injection of 
impacting objects into the terrestrial planet zone accounts for 
the abrupt end of the intense bombardment; thus, this event has 
been named the Late Heavy Bombardment (LHB), or sometimes the 
Lunar Cataclysm, to distinguish it from the prior final accretion 
of the planets at 4.5 Ga. Specifically, the lunar cataclysm 
hypothesis {\it (1,2)} postulates that the intense bombardment of the 
Moon lasted only a very short period of time, 20--200 My {\it (2--5)}. 
Recent results on the impact ages of lunar meteorites (which 
represent a much broader region of the lunar surface than the Apollo 
samples) support this hypothesis {\it (6--8)}. Furthermore, the 
impact-reset ages of meteoritic samples of asteroids {\it (9--10)}
and the shock-metamorphosing at 3.92 Ga of the only known sample of 
the heavily cratered highlands of Mars, meteorite Allan Hills 84001 
{\it (11)}, indicate that the LHB affected the entire inner Solar System, 
not just the Moon. 

Identifying the sources of planetary impactors has proven elusive. 
Dynamical models invoking both geocentric and heliocentric debris 
and both asteroidal and cometary reservoirs have been proposed {\it (12)}, 
but chemical analyses of Apollo samples of impact melts point to a 
dominantly asteroid reservoir for the lunar cataclysm {\it (10)}. In this 
paper, we provide compelling new evidence that the source of the LHB 
impactors was the main asteroid belt, and that the dynamical mechanism 
that caused the LHB was unique in the history of the Solar System and 
distinct from the processes producing the flux of objects that currently 
hit planetary surfaces. 

We examined the crater size distributions {\it (13)} of surface units of 
various ages on the Moon, Mars and Mercury, based on published data 
{\it (14,15)} supplemented by new crater counts {\it (S1)}. Of the 
terrestrial planets, only the Moon, Mercury and Mars have heavily 
cratered surfaces. These surfaces all have complex size distributions, 
as seen in Fig.~1a. The curves for Mercury and Mars are steeper than the 
lunar curve at diameters less than about 40 km, because plains formation 
has obliterated a fraction of the smaller craters {\it (S2)}. Therefore, the 
lunar highlands curve best represents the shape of what we shall call 
the Population 1 crater size distribution. 

In addition to the highlands, the crater curves for Martian old plains 
east of the Tharsis region, old plains within the Hellas basin and 
plains within and surrounding the Caloris basin on Mercury also have 
the same shape as the lunar highland curve over the same diameter range, 
but with a lower crater density {\it (16)}. The lower crater densities imply 
that these older plains probably formed near the tail end of the LHB, 
about 3.8 billion years ago.

For the younger surfaces, the crater size distribution curves are flat 
and distinctly different (Fig.~1b). These include the lightly cratered, 
hence younger, plains on Mars and the Moon, as well as fresh craters 
with well-defined ejecta blankets (Class 1 craters) on the Moon.  
This crater population we call Population 2.

The crater density on Venus (Fig.~2) is about an order of magnitude 
less than on Mars. Only young craters are present, evidently 
because older craters have been erased by multiple global resurfacing 
events {\it (17)}. Furthermore, small craters are scarce on Venus because its 
thick atmosphere screens out small impactors {\it (18)}. Part of the Venus 
crater population consists of clusters of craters (multiples) that 
result from fragmentation of the impacting object in the dense 
atmosphere. These comprise 16\% of all Venus craters {\it (S1)}. The size 
distribution of these multiples is also shown in Fig.~2, where the 
diameter is derived from the sum of the crater areas in the cluster. 
The turnover of the curve for multiple craters does not occur until 
diameters less than 9 km; at larger diameters the curve is flat. This, 
together with the much lower crater density, strongly suggests that the 
impacting population on Venus was the same as Population 2 on the Moon 
and Mars. It is also evidence that the turnover of the crater curve is 
indeed due to atmospheric screening.

The two characteristic shapes of the crater curves in the inner Solar 
System are summarized in Fig.~3. We conclude that the terrestrial planets 
have been impacted by two populations of objects which are 
distinguishable by their size distributions. Population 1 is responsible 
for the LHB, and Population 2 is responsible for the post--LHB period and 
up to the present time.

A number of studies on the physics of impact cratering on solid bodies 
have derived projectile--crater scaling laws. We used the Pi scaling law 
{\it (19)} to derive the projectile size distribution for Population 1 and 
Population 2 impactors.  We used the lunar highland crater curves as 
representative of Population 1 and the Martian young plains as 
representative of Population 2, as these provide the best statistics. 
(We did not include crater diameters greater than 500 km, because of 
scaling uncertainties.)  We assumed projectile parameters appropriate for 
asteroidal impacts: density of 3 g cm$^{-3}$ (similar to basaltic rock), an 
impact angle of 45$^\circ$ and impact velocities of 17 km s$^{-1}$ and 
12 km s$^{-1}$ on the Moon and on Mars, respectively {\it (20)}. 
In Fig.~4, we compare these distributions to recent determinations of 
the size distributions of the main belt asteroids (MBAs) {\it (21--23)} and 
Near Earth Asteroids (NEAs) {\it (24)}. The size distribution of the current 
MBAs is virtually identical to the Population 1 projectile size 
distribution, as pointed out in {\it (25)}.  This result indicates that the 
objects responsible for the LHB originated from Main Belt asteroids. 
Unless comets or Kuiper belt objects have the same size distribution, 
these objects could not have been significant contributors to the LHB.

The close match between the current MBA size distribution and that of the 
LHB projectiles implies that the main asteroid belt has remained unchanged 
in its size distribution over the past 3.8 Gy. There are two possible 
interpretations of this result: either collisional processes produced a 
steady-state size distribution in the main asteroid belt at least as early 
as 3.8 Ga, or the collision frequency in the main asteroid belt was 
drastically reduced around 3.8 Ga. 

The mechanism responsible for ejecting asteroids from the main 
asteroid belt and into terrestrial planet-crossing orbits during 
the LHB had to be unique to the early Solar System because there 
is no evidence for any subsequent event of similar magnitude in 
the inner planets' cratering history since then. Furthermore, 
that mechanism had to be one that ejected asteroids from the main 
belt in a size--independent manner, preserving the MBA size 
distribution in the inner planet impactor population. This precludes 
size--dependent non--gravitational transport processes, 
such as the Yarkovsky effect, 
and instead implicates a dynamical process, such 
as sweeping gravitational resonances, that was largely insensitive 
to asteroid mass. 

A dynamical mechanism consistent with the above constraints that explains 
the congruence of the size distributions of the MBAs and the Population 1 
projectiles, involves the orbital migration of the giant planets.  
Such migration of the outer planets is thought to have occurred on a 
timescale of about $10^7$--$10^8$ years early in Solar System history 
{\it (26--29)}, and it would have caused severe depletion of the asteroid 
belt due to orbital instabilities that ensue as strong Jovian mean motion 
resonances sweep across this zone {\it (30)}. This phenomenon would have 
caused the Moon and terrestrial planets to be cataclysmically bombarded by 
asteroids and icy planetesimals (comets) for a period of 10--100 Myr 
{\it (31)}.  
A recently proposed variation on the giant planet migration theory invokes 
the change of the eccentricities of Jupiter and Saturn, if and when these 
planets passed through a 1:2 orbital resonance during their orbital 
migration {\it (32)}. Such resonance passage would have destabilized the 
planetesimal disk beyond the orbits of the planets, causing a sudden massive 
delivery of comets to the inner Solar System; in this scenario,
the asteroid belt is also destabilized due to sweeping gravitational
resonances; together, these cause a significant spike in the intensity of 
cometary as well as asteroid impacts on the inner planets {\it (33)}. 

In either scenario, the relative intensity of comets versus asteroids in the 
projectile population of the LHB is not well determined by the published 
dynamical simulations.  Since the impact signature in the crater record in 
the inner Solar System is asteroidal, we conclude that either comets played 
a minor role or their impact record was erased by later--impacting asteroids.

Both of these mechanisms predict a LHB lasting between $\sim\!10$ My and 
$\sim\!150$ My. Therefore, the LHB was a catastrophic event that occurred 
from about 3.9 Ga to 3.8 Ga. Because of this, it is not possible to use the 
crater record to date surfaces older than about 3.9 Gy; 
the previous crater record has been obliterated by this event. The heavily 
cratered highlands of the Moon, Mars and Mercury that register Population 1 
were resurfaced 3.9 billion years old, although older rock relics may have 
survived. 

The size distribution of Population 2 projectiles (Fig.~4) is the 
same as the NEAs and quite different from the LHB projectiles. Thus, 
NEAs are largely responsible for the cratering record after 3.8 Ga. 
This result is contrary to the findings in {\it (34)} that apparently used 
biased data containing observational losses (cf.~ref.~24) and whose analysis 
based on cumulative (rather than differential) size distributions was
not sufficiently sensitive to the differences in the distributions.

A plausible reason that the MBAs and the NEAs have such a different 
size distribution is the Yarkovsky effect that causes secular changes 
in orbital energy of an asteroid due to the asymmetric way a spinning 
asteroid absorbs and reradiates solar energy {\it (35)}. Over a few 
tens of millions of years the effect is large enough to transport 
a significant number of sub--20 km size asteroids into strong 
Jovian resonances {\it (36)} that deliver them into terrestrial 
planet-crossing orbits.  The magnitude of the effect depends on the 
size of the asteroid: for diameters greater than about 10 m, 
the smaller the asteroid the larger the effect. This explains why 
the NEAs (Population 2 projectiles) have relatively more small 
objects compared to MBAs. Because the younger post-LHB surfaces have 
been impacted primarily by NEAs, the ages of these surfaces can be 
derived from the crater production rate of NEAs. However, the ages 
derived from the NEA impacts will be an upper limit because we do 
not know the comet crater production rate with any certainty.

Our results further imply that dating surfaces of solid bodies in 
the outer solar system using the inner planet cratering record is 
not valid. Attempts have been made to date outer planet surfaces on 
an absolute time scale by assuming that the crater population found 
in the inner Solar System is the same throughout the entire 
Solar System and has the same origin. In light of our results, 
this assumption is false. Additional evidence to support this 
conclusion is found in the cratering record of the Jovian satellites. 
Indeed, Callisto has a crater size distribution different than both 
Population 1 and Population 2 craters {\it (37,38)}.

\newpage

\noindent{\bf Figure Captions}
\bigskip

\noindent{\bf Fig.~1.} The crater size distributions on the Moon, Mars and 
Mercury, shown as R plots {\it (13)}.  The upper curves (a) are for heavily 
cratered surfaces on the Moon (blue), Mars (red) and Mercury (green). The 
lower curves (b) are for younger surface on the Moon (blue) and Mars (red).  
The size distributions on younger surfaces (Population 2) are different than 
for the old surfaces that represent the LHB (Population 1).

\noindent{\bf Fig.~2.} Size distributions of all Venus craters and, 
separately, multiple craters, compared to the Mars Northern Plains (green). 
The downturn in the Venus curves (dashed blue lines) is due to atmospheric 
screening of projectiles. The unscreened portions (red) are the same as 
Population 2 on Mars.

\noindent{\bf Fig.~3.} These crater curves summarize the inner Solar System 
cratering record, with two distinctly different size distributions. The red 
curves are Population 1 craters that represent the period of Late Heavy 
Bombardment. The lower density blue curves (Population 2) represent the 
post-LHB era on the Moon, Mars and Venus. The Mars Young Plains curve is a 
combination of the the Mars Northern Plains and Mars Young Volcanics.  The 
Venus curve is a composite of the production population for all craters 
greater than 9 km, including multiples in the range of 9--25 km diameter.

\noindent{\bf Fig.~4.} The size distributions of the projectiles (derived 
from the crater size distributions) compared with those of the Main Belt 
Asteroids (MBAs) and Near Earth Asteroids (NEAs).  The red dots (upper curve)
are for the lunar highlands (Population 1), and the red squares (lower curve)
are for the Martian young plains (Population 2). 
The other colors and point styles are for the asteroids derived by various 
authors: in the upper curves, the light blue, the dark blue and the green 
symbols are from Spacewatch {\it (21)}, the Sloan Digital Sky Survey 
{\it (22)}, and from the Subaru asteroid surveys {\it (23)}, respectively; 
the black dots in the lower curves are the debiased LINEAR near earth 
asteroids {\it (24)}.  An arbitrary normalisation factor is applied to 
obtain the R values for the asteroids.
The MBA size distribution is virtually identical with Population 1 
projectiles responsible for the LHB crater record. The NEA size 
distribution is the same as Population 2 projectiles responsible for 
the post--LHB crater record.

\newpage
\begin{figure}
\includegraphics*[scale=0.7]{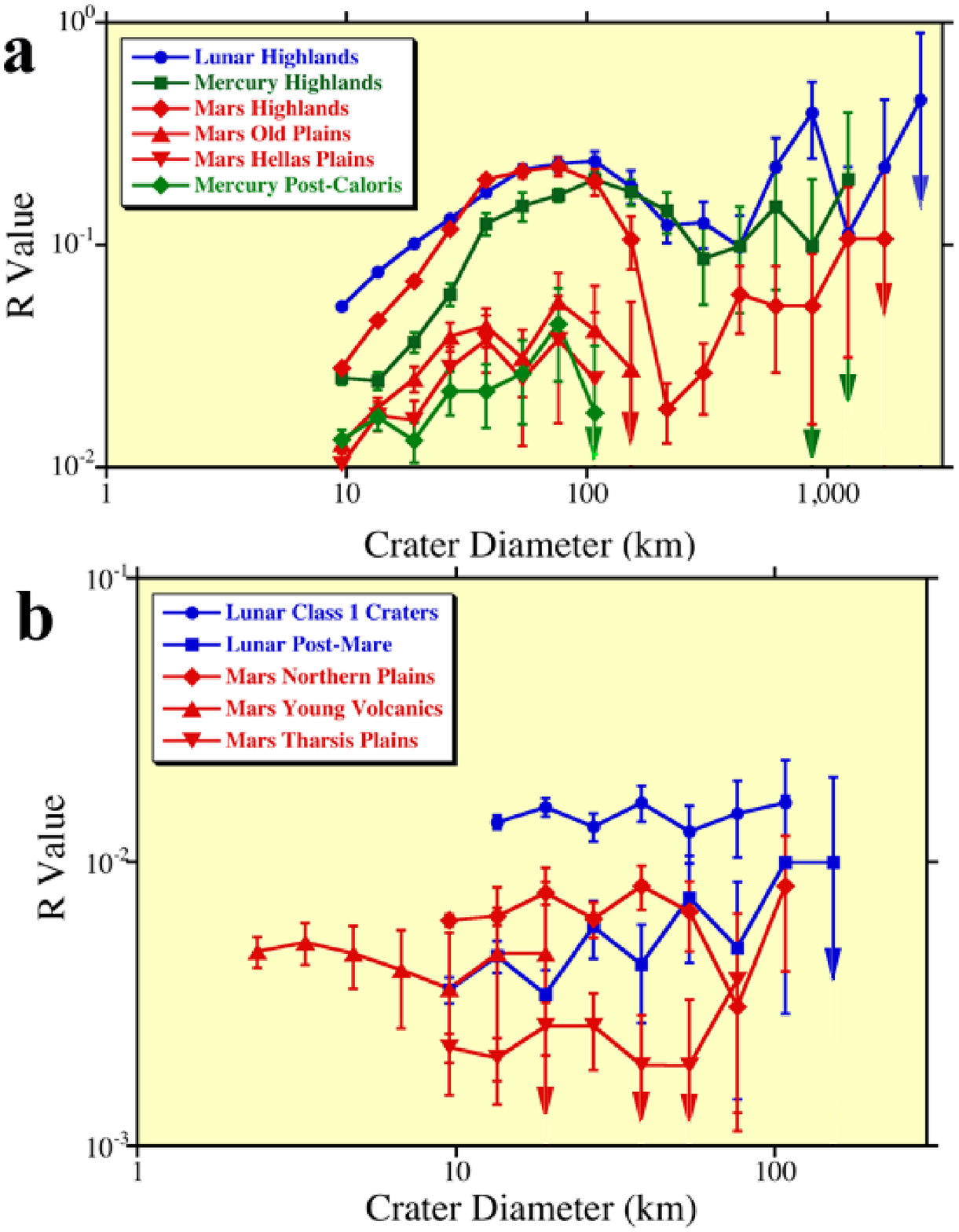} \label{fig-1}
\end{figure}

\vfill\eject

\begin{figure}
\includegraphics[scale=0.7]{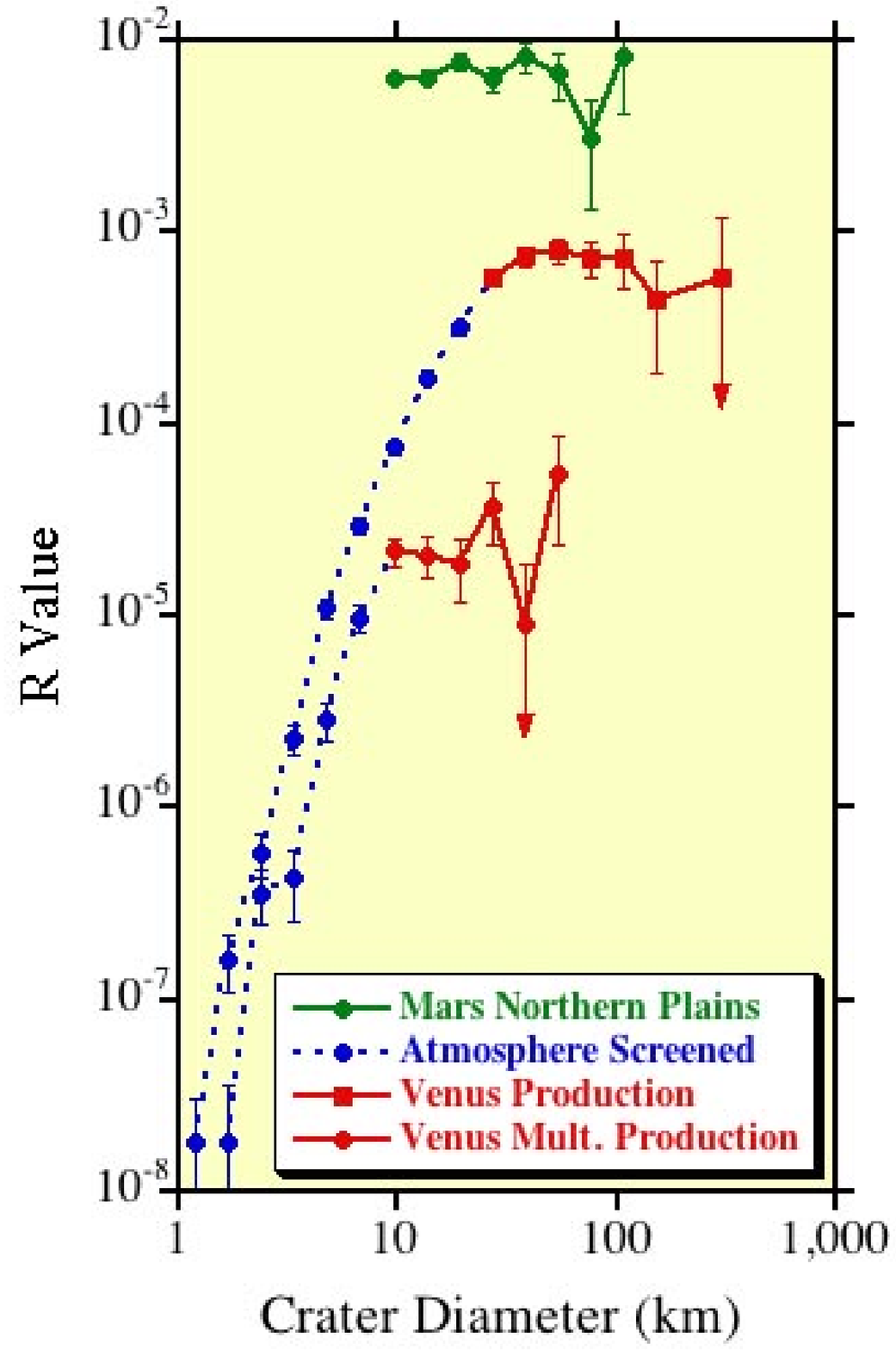} \label{fig-2}
\end{figure}

\vfill\eject

\begin{figure}
\includegraphics[scale=0.7]{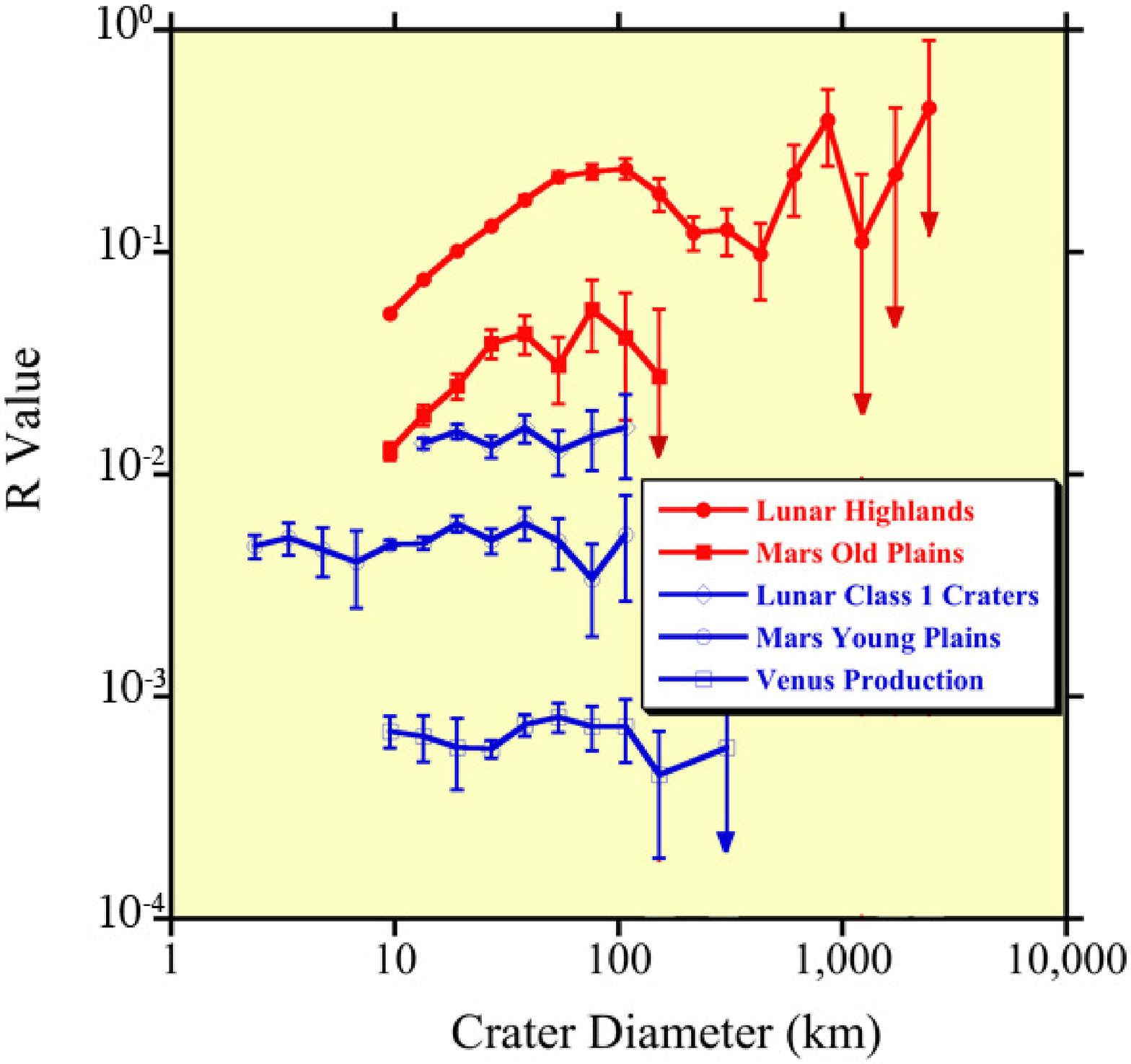} \label{fig-3}
\end{figure}

\vfill\eject

\begin{figure}
\includegraphics*[scale=0.7]{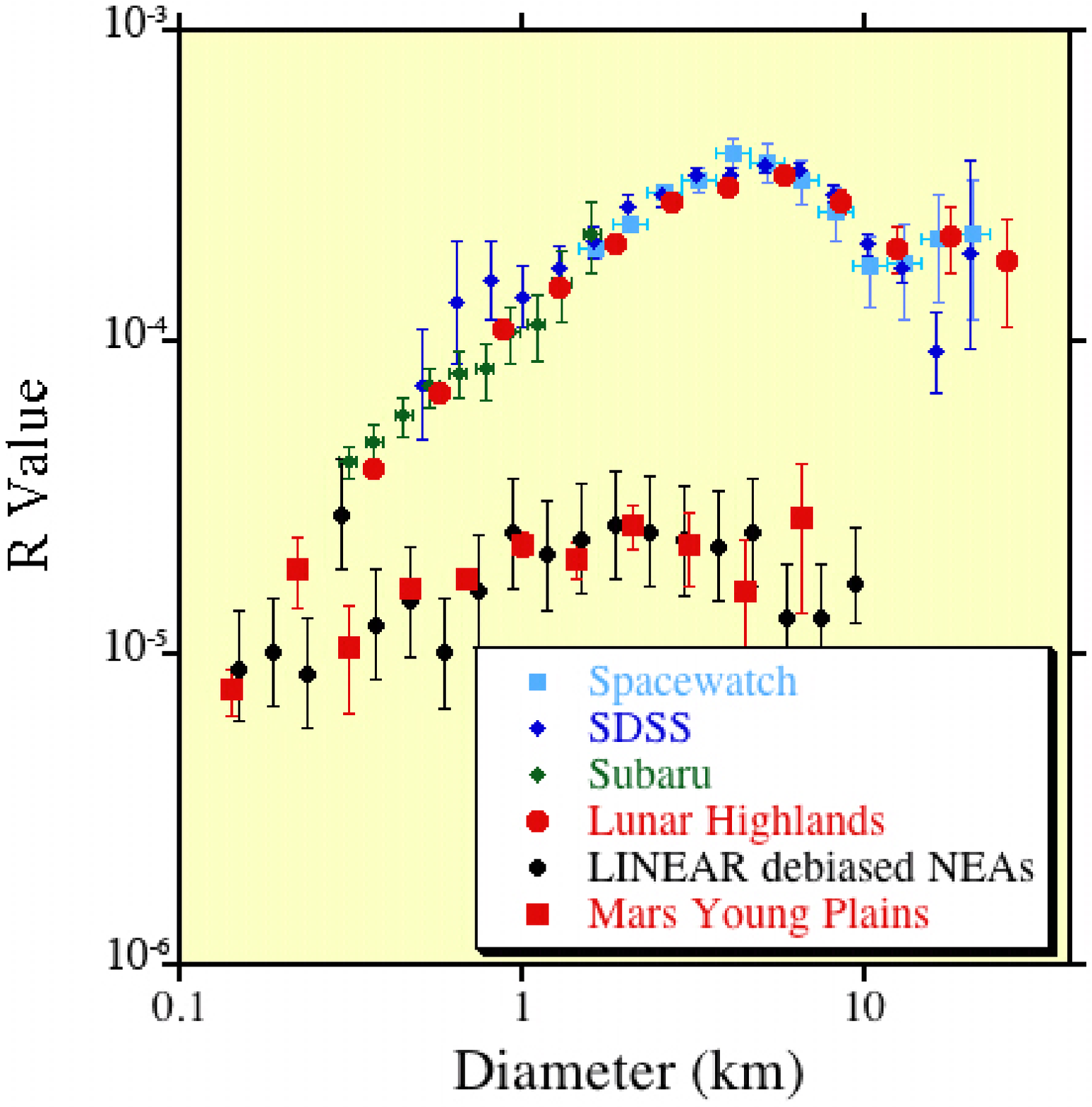} \label{fig-4}
\end{figure}


\begin{thebibliography}{}
\bibitem[cratersaturation (23)]{cratersaturation}
16. This also demonstrates that the shape of the lunar highlands curve has not 
been affected by crater saturation because they show the same size 
distribution as the highlands but the crater densities are well below 
saturation density, confirming a theoretical result that a surface impacted 
by a population with the same size distribution as the one observed for the 
highlands would maintain the same shape at saturation  (Woronow, A., 
\jgr, 82, 2447-2451, 1977).
\end{thebibliography}
\end{document}